\documentclass[12pt,aps,showpacs]{revtex4}%
\usepackage{amssymb}
\usepackage{amsmath}
\usepackage{amsfonts}
\usepackage{graphicx}%
\providecommand{\U}[1]{\protect\rule{.1in}{.1in}}
\providecommand{\U}[1]{\protect\rule{.1in}{.1in}}
\providecommand{\U}[1]{\protect\rule{.1in}{.1in}}
\providecommand{\U}[1]{\protect\rule{.1in}{.1in}}
\providecommand{\U}[1]{\protect\rule{.1in}{.1in}}
\providecommand{\U}[1]{\protect\rule{.1in}{.1in}}
\providecommand{\U}[1]{\protect\rule{.1in}{.1in}}
\providecommand{\U}[1]{\protect\rule{.1in}{.1in}}
\providecommand{\U}[1]{\protect\rule{.1in}{.1in}}
\begin{document}
\title{Thin Shell Dynamics and Equations of State}
\author{J.P. Krisch and E.N. Glass}
\affiliation{Department of Physics, University of Michigan, Ann Arbor, MI 48109}
\date{July 7, 2008}

\begin{abstract}
A relation betweeen stress-energy and motion is derived for accelerated Israel
layers. The relation, for layers between two Schwarzschild manifolds,
generalizes the equation of state for geodesic collapse. A set of linked
layers is discussed. \ 

\end{abstract}

\maketitle

\section{Introduction}

Equations of state and boundary matching are two important tools for
developing exact solutions of Einstein's field equations. As models have
become more physical, the requirements of strict boundary matching have been
relaxed by joining two exact solutions across a boundary layer, matching their
metrics on the layer but allowing jumps in the derivative structure.\ The
Israel junction conditions \cite{Isr66} are often used to make the broader
matching and are widely applied because they provide a simple dynamic boundary
description for a variety of scenarios ranging from thin shell descriptions
\cite{KL79,KM91,MK96,Gon02}, to shell applications like bubbles
\cite{LW86,L-CM86,BKT87,Ips87,KMM06}, walls \cite{IS84, Ips84, GV89},
gravastars \cite{MM01, VW04, Car05, CFV05, deBHI+06}, and extensions of
general relativity like dilatons \cite{ES05} or Gauss-Bonnet gravity
\cite{BI05,GW07}. Since their introduction, Israel layers \cite{Isr66, Bar91}
have played an increasingly important role in gravitational physics. Barrabes
and Israel \cite{Bar91} began their paper with a description of the Israel
layer as a thermodynamic phase boundary, but the initial applications of
Israel layers considered metric matching in dynamic collapse processes
involving dust shells, null shells, and cosmic string loops \cite{Isr66,
Bar91}. Poisson \cite{Poi04} has summarized some of the early seminal work
\cite{Isr66,Bar91,CI68,ML97}. As our knowledge of the variety of astrophysical
objects and their dynamic processes has expanded, Israel layers have become
physically interesting in their own right and the questions that have been
investigated for large scale three-dimensional mass distributions, are now
being asked about Israel layers \cite{BLP91,Ver02,Ans02,LC05,Mar05}.

In this paper we investigate the relation between a layer's equation of state
and its motion.\ An equation of state for a layer dropping from rest at
infinity on an exterior Schwarzschild geodesic has been formulated
\cite{KG07}, and can be generalized to include accelerated motions. We develop
the extension and apply it to static and dynamic examples. The motion input to
the geodesic extension is a single function.\textbf{ \ }In the next section we
briefly review the thin shell formalism used in the rest of the paper and give
the geodesic extension. The applications of the extension are in Section III.
Many of the applications use a layer with a linear equation of state,
$P=a\sigma$ (L-layer). The description of an L-layer has been included in
Appendix B. \ The accelerated extension singles out several special $^{\prime
}a^{\prime}$ values, $a=(-1/2,-1/4,0)$.\ We show that the $a=-1/4$ L-layer is
simply related to a geodesic layer. A set of linked L-layers is described and
the other two special $^{\prime}a^{\prime}$ values are shown to be boundaries
for the linked layers. \ \ 

\section{Thin Shell Description}

We consider an Israel layer as a thin shell $\Sigma$ between two Schwarzschild
manifolds with exterior mass parameter $m_{0}$ and interior mass parameter
$M$. The spacetime consists of the two manifolds which join across surface
$\Sigma$ (+/- denotes exterior/interior).%
\begin{subequations}
\begin{align}
g_{ab}^{\pm}dx^{a}dx^{b}  &  =-(f_{\pm})dt^{2}+(1/f_{\pm})dr^{2}+r^{2}%
d\Omega^{2},\\
f_{+}  &  =1-2m_{0}/R,\text{ \ }f_{-}=1-2M/R,\\
ds_{\Sigma}^{2}  &  =-d\tau^{2}+R^{2}(\tau)d\Omega^{2}.
\end{align}
The layer is tracked by two observers comoving with the layer. The observers
use $r=R(\tau)$ and $T_{\pm}=T_{\pm}(\tau)$ to describe the layers. With this
parametrization and with overdots denoting $d/d\tau$, the observers'
velocities and associated normal vectors are%
\end{subequations}
\begin{align}
U_{\pm}^{i}  &  =(\dot{T}_{\pm},\dot{R},0,0),\text{ \ }i=t,r,\vartheta
,\varphi\\
n_{i\pm}  &  =(-\dot{R},\dot{T}_{\pm},0,0).
\end{align}
Velocity normalization $g_{ij}^{\pm}U_{\pm}^{i}U_{\pm}^{j}=-1$ implies
$(f_{\pm})^{2}\dot{T}_{\pm}^{2}=\dot{R}^{2}+f_{\pm}.$ The jump in $f_{\pm}$,
$\dot{T}_{\pm}$, is an important input to layer motion, and we define%
\begin{align}
\Delta &  =\Delta_{+}-\Delta_{-}\label{delta}\\
\Delta_{\pm}  &  =f_{\pm}\dot{T}_{\pm}=\sqrt{\dot{R}^{2}+f_{\pm}}.
\end{align}

\subsection*{Matter Content}

The layer has stress-energy content%
\begin{equation}
S_{\ j}^{i}:=\sigma U^{i}U_{j}+P(g_{\ j}^{i}+U^{i}U_{j})
\end{equation}
with density $\sigma$ and stress $P$ ( $-P$ is tension). The stress-energy
content of layers depends on the bounding metrics and the layer motion. From
the Israel conditions \cite{Poi04} we have%
\begin{align}
4\pi\sigma &  =-\Delta/R,\label{sigma}\\
8\pi P  &  =\Delta/R+\dot{\Delta}/\dot{R}, \label{p}%
\end{align}
with the layer mass $m_{L}$ defined by
\begin{equation}
m_{L}=4\pi R^{2}\sigma.
\end{equation}

\subsection*{Layer Motion}

Motion parameters $\dot{R}^{2}$ and $\ddot{R}$ are both important inputs to
the stress-energy structure of the layer.\
\begin{align}
\dot{R}^{2}  &  =\left(  \frac{\Delta}{2}\right)  ^{2}+\left(  \frac{m_{0}%
-M}{R\Delta}\right)  ^{2}+\frac{m_{0}+M}{R}-1,\label{rd2}\\
\ddot{R}  &  =\frac{\dot{\Delta}}{\dot{R}}\left[  \frac{\Delta}{4}%
-\frac{(m_{0}-M)^{2}}{\Delta^{3}R^{2}}\right]  -\frac{m_{0}+M}{2R^{2}}%
-\frac{(m_{0}-M)^{2}}{\Delta^{2}R^{3}}.
\end{align}
The radial components of the 4-accelerations of observers comoving with the
layer (Appendix A) are%
\begin{equation}
\dot{U}_{\pm}^{r}=\ddot{R}+(1-f_{\pm})/(2R).
\end{equation}
The radial accelerations are related to a stress-energy sum%
\begin{equation}
4\pi(\sigma+2P)=\frac{\dot{\Delta}}{\dot{R}}=\frac{\dot{U}_{+}^{r}}%
{\sqrt{1+\dot{R}^{2}-2m_{0}/R}}-\frac{\dot{U}_{-}^{r}}{\sqrt{1+\dot{R}%
^{2}-2M/R}}. \label{u-diff}%
\end{equation}
These relations provide insight about the role of jumps in the observer
accelerations and velocities in defining the stress-energy. It is clear that
the stress-energy structure of the layer and its motion are related so that
different assumptions about the motion will produce different equations of
state or inversely, that an imposed equation of state will determine the layer
motions. \ 

\subsection*{The Geodesic Generalization}

A layer starting from rest at infinity and collapsing along a geodesic in the
exterior spacetime has motion parameters%
\begin{subequations}
\begin{align}
\dot{R}_{g}  &  =-\sqrt{2m_{0}/R},\\
\ddot{R}_{g}  &  =-m_{0}/R^{2},\\
R_{g}(\tau)  &  =(2m_{0})^{1/3}[(3/2)(c_{1}-\tau)]^{2/3}.
\end{align}
The equation of state for the geodesic layer \cite{KG07}\ is%
\end{subequations}
\begin{equation}
\sigma(1+4P/\sigma)^{3}=(2P/\sigma)^{2}\ \frac{(1+2P/\sigma)}{\pi(m_{0}-M)}.
\end{equation}
This geodesic equation of state is the limit ($\gamma\rightarrow0,$
$A\rightarrow0)$ of a layer with exterior motion parameters
\begin{align}
\gamma &  =\dot{R}^{2}-\frac{2m_{0}}{R},\\
A  &  =\dot{U}_{+}^{r}=\ddot{R}+\frac{m_{0}}{R^{2}},
\end{align}
where $A$ is the radial 4-acceleration of the exterior comoving observer and
$1+\gamma=\Delta_{+}^{2}.$ $A$ and $\gamma$ are related by $A=\dot{\gamma
}/(2\dot{R})$. With these parameters, Eq.(\ref{u-diff}) can be written as%
\begin{equation}
4\pi(\sigma+2P)=\frac{A}{\sqrt{1+\gamma}}-\frac{A+(M-m_{0})/R}{\sqrt
{1+\gamma+2(m_{0}-M)/R}}. \label{a-diff}%
\end{equation}
In this equation, the motion parameters, $R,$ $A,$ and $\gamma$ require
several separate choices to set integration constants. The
stress-energy-motion relation (Appendix C ) involving a single motion function
$F:=AR/(1+\gamma)$ is
\begin{equation}
\pi\sigma(m_{0}-M)(1+4P/\sigma)^{3}=(1+\gamma)^{3/2}\left(  F-2P/\sigma
\right)  ^{2}(1+F+2P/\sigma). \label{sig-cubed}%
\end{equation}
This is the generalization of the geodesic equation of state. For $m_{0}=M$
there is no layer and the motion is described by $\dot{R}A=0$, $\gamma=const$.
In developing applications of this equation, one notes that two key inputs are
either the ratio $P/\sigma$ or the value of $F.$ Choosing a static layer
$(\dot{R}=0,$ $\ddot{R}=0)$ sets the value of $F$ and a general equation of
state results. Choosing $P/\sigma$ selects the motion; for example, the
equation of state $P/\sigma=(-1/4,-1/2,0)$ are all simplifying special values
for the relation.\ In the next section we will examine all three values.\ We
begin with dynamic layers and consider static layers as a second example. \ 

\section{Application to non-geodesic layer motion}

\subsection{Dynamic Layers: P=-(1/4)$\sigma$}

$P=a\sigma=-(1/4)\sigma$ is singled out by the generalized
stress-energy-motion relation. From Eq.(\ref{sig-cubed}), for this equation of
state, one has%
\begin{equation}
F=\frac{AR}{1+\gamma}=-1/2. \label{ar-half}%
\end{equation}
This can be integrated for $1+\gamma$ giving%
\begin{equation}
1+\gamma=\frac{C_{0}}{R}%
\end{equation}
with $C_{0}$ an integration constant. Using the motion function, $A$ is
\begin{equation}
A=-\frac{C_{0}}{2R^{2}}=\frac{C_{0}}{2m_{0}}\ \ddot{R}_{g}.
\end{equation}
Layers that have a radial 4-acceleration linearly related to the geodesic
$\ddot{R}_{g}$ value will have the $P=-\sigma/4$ equation of state.\ They are
L-layers with tension. In the section on static layers, we will see that
$a=-1/4$ is an excluded value for static L-layers. We will also see that
$a=-1/4$ is an important boundary point for some of them. Note that there are
moving L-layers with tension.

\subsection{Linked dynamic layers}

Because of its simplicity, the linear equation of state is often used in
discussing thin shells \cite{LW86,GV89,NU201}.\ For L-layers (Appendix B)%
\[
\Delta=-c_{a}R^{-(1+2a)}.
\]
For this $\Delta$, the motion of the layer is described by Eq.(\ref{rd2})
\begin{equation}
\dot{R}^{2}=\frac{c_{a}^{2}R^{-2(1+2a)}}{4}+\frac{(m_{0}-M)^{2}}{c_{a}%
^{2}R^{-4a}}+\frac{m_{0}+M}{R}-1. \label{dot-sq}%
\end{equation}
There is an interesting symmetry in this equation. Over the positive
$^{\prime}a^{\prime}$ range, $a=a_{p}$, $0\leq a_{p}\leq1$ Eq.(\ref{dot-sq})
describes the motion. For$\ a=-a_{n},$ $0\leq a_{n}\leq1$ the motion is
described by%
\begin{equation}
\dot{R}^{2}=\frac{c_{a}^{2}R^{-2(1-2a_{n})}}{4}+\frac{(m_{0}-M)^{2}}{c_{a}%
^{2}R^{4a_{n}}}+\frac{m_{0}+M}{R}-1.
\end{equation}
The substitutions
\begin{align}
a_{n}  &  =a_{p}+1/2,\label{a-n}\\
c_{a_{n}}^{2}  &  =4(m_{0}-M)^{2}/c_{a_{p}}^{2},\nonumber
\end{align}
map these two motion equations into each other.\ The $^{\prime}a^{\prime}$
range over which a layer with tension is linked to a layer with pressure is%
\begin{align}
1/2  &  \leq a_{n}\leq1\\
0  &  \leq a_{p}\leq1/2.\nonumber
\end{align}
For each negative $a=-a_{n}$ in this range, there is a layer with tension
which has the same motion as a layer with pressure and positive $a=a_{p}$. The
linked layers have the same $\dot{R}$ and $\ddot{R}$.\ $a_{p}=0$ and
$a_{n}=1/2$ are the lowest\ $^{\prime}a^{\prime}$ values for linked
pressure/tension shells and are two of the values which simplify
Eq.(\ref{sig-cubed}). For these two values we have%
\begin{align}
a  &  =0:\ \ \ \ \ \ \pi(m_{0}-M)\sigma_{a=0}=(1+\gamma)^{3/2}F^{2}%
(1+F)\label{a-zero}\\
a  &  =-1/2:\ \ \pi(m_{0}-M)\sigma_{a=-1/2}\ =-(1+\gamma)^{3/2}(F+1)^{2}(F)
\label{a-one}%
\end{align}
The motion functions are%
\begin{align*}
a  &  =0:\ \ \ \ \ \ \ F=-\frac{c_{0}^{2}}{R[c_{0}^{2}-2R(m_{0}-M)]},\\
a  &  =-1/2:\ \ F=-\frac{2(m_{0}-M)}{R[2(m_{0}-M)-Rc_{-1/2}^{2}]},
\end{align*}
and are identical under the linkage. The matter content of the two layers is
different. From Eq.(\ref{a-zero}) and Eq.(\ref{a-one}) we have%
\[
\frac{\sigma_{a=0}}{\sigma_{a=-1/2}}=-\frac{F}{1+F}.
\]
For $m_{0}=M$, we would expect no layer to exist and $a=-1/2$ describes
Schwarzschild vacuum.

A missing part of the $^{\prime}a^{\prime}$ range for linked pressure/tension
shells is $-1/2<a<0$. There are linked L-layers in this region but the linked
layers both have tension. This linkage is centered around $a=-1/4$, one of the
special values for the geodesic extension, describing layers whose radial
4-accelerations are linearly related to $\ddot{R}_{g}$.\ Consider two negative
$^{\prime}a^{\prime}$ values, $a_{n_{1}}$ and $a_{n_{2}}$ in the ranges%
\begin{align}
1/2  &  <a_{n_{1}}\leq1/4\\
1/4  &  \leq a_{n_{2}}<0.\nonumber
\end{align}
As might be expected from the pressure/tension linking relations, the relation
for these layers are%
\begin{align}
a_{n_{1}}  &  =-a_{n_{2}}+1/2\\
c_{a_{n1}}^{2}  &  =4(m_{0}-M)^{2}/c_{a_{n2}}^{2}.
\end{align}
When $a=-1/4,$ then $a_{n_{1}}$ and $a_{n_{2}}$ coincide.

\subsection{Static layers}

Static layers with $\dot{R}=0,$ $\ddot{R}=0$ have motion parameters%
\begin{align}
\gamma_{s}  &  =-2m_{0}/R_{s},\\
A_{s}  &  =m_{0}/R_{s}^{2},\nonumber
\end{align}
and are, in some sense, the negatives of the geodesic layer with $\gamma
_{s}=-\dot{R}_{g}^{2},$ $A_{s}=-\ddot{R}_{g}.$ Using Eq.(\ref{p}) and
Eq.(\ref{sig-cubed}) we have
\begin{equation}
4P/\sigma=-1+\frac{1}{1-2M/R_{s}-4\pi R_{s}\sigma\sqrt{1-2M/R_{s}}}.
\label{big-denom}%
\end{equation}
For $M=0$ this is the stress energy-radius relation for static layers given by
Khourami and Mansouri\ \cite{KM91}, with their stress and density related to
$P$ and $\sigma$ by $P_{km}=8\pi P,$ $\sigma_{km}=8\pi\sigma$
\begin{equation}
P=\frac{\pi R_{s}\sigma^{2}}{1-4\pi R_{s}\sigma}. \label{stat-eos}%
\end{equation}
This relation is particularly interesting when compared with the classical van
der Waals form for 3-dimensional fluids.%
\[
P=\frac{nRT}{V-nb}-a(\frac{n}{V})^{2}%
\]
with $n$ the number of moles, volume $V$, gas constant $R$, and temperature
$T$. $a$ and $b$ are constants.\ The denominator of the first term corrects
for a minimum volume available to the fluid constituents.\ This can be
attributed to a finite size of the particle constituents or to the existence
of a repulsive core in constituent interactions.\ The second term accounts for
an attractive long range attraction between constituents which reduces the
stress. For low densities ($n/V$) the equation of state describes a perfect
fluid, $PV=nRT$.

The numerator of Eq.(\ref{stat-eos}) could imply that, for $\pi R_{s}%
\sigma<<1$, an Israel layer is a first order polytrope. The current
relativistic polytrope assumes a linear low density equation of
state\ \cite{Lob06}.\ The part of the classical van der Waals equation
describing long range interaction is missing\ from Eq.(\ref{stat-eos}) but the
denominator suggests, as in the classical equation, there is either a minimum
or a zero.\ The existence of a minimum value could be related to the existence
of a repulsive core in the interaction potential between allowed layer
constituents.\ Detailed models of this possibility will be discussed
elsewhere.\ Equation (\ref{stat-eos}) also suggests that there are static
shells with tension and pressure with the zero denominator related to the
boundary between the two kinds of stress.\ At the zero value the Israel
$\Delta$ can be found from the density%
\begin{align}
4\pi R_{s}\sigma &  =1,\nonumber\\
\Delta &  =-1.\nonumber
\end{align}
$\dot{R}=0$ is one of the static shell conditions.\ From the definition of
$\Delta$, Eq.(\ref{delta}), we have for an $M=0$ static shell \
\[
\Delta=\sqrt{1-\frac{2m_{0}}{R}}-1.
\]
We see that there are no static shells outside the horizon corresponding to
$\Delta=-1.$ The boundary between the static linked layers is not static.

The static shell boundaries can be explored using\ $M=0,$ L-shells as an
example. For static L-shells the Israel radius value is (Appendix B)
\[
R_{s}=m_{0}\frac{(4a+1)^{2}}{4a(1+2a)}%
\]
and the range for physical static radii is $-1\leq a<1/2,$ $0<a\leq1$. There
are linked static shells with tension and pressure, just as in the dynamic
case. For each $R_{s}$ in the range $0<a\leq1/2,$ there is an identical radius
in the range \ $-1\leq a<-1/2.$ For example, $a=1/2$ and\textbf{ }$a=-1$ have
the same static radius $R_{s}=\frac{9}{4}m_{0}$, as suggested by the linking
relation $a_{n}=a_{p}+1/2$. The range for static shells excludes the points
$a=-1/2$ and $a=0$. These points are the moving boundaries of the static layer
region. The boundary between the pressure/tension linked $M=0$ static L-layers
is a moving layer that is not the dust layer one might have expected, but the
layer linked to dust. \ \ 

The $M\neq0$ static relation is less easy to interpret in a van der Waals
sense, since the size of the interior mass $M$ becomes an important
parameter.
\begin{equation}
P=\frac{\pi R_{s}\sigma^{2}}{\sqrt{1-2M/R_{s}}-4\pi R_{s}\sigma}+\frac
{M\sigma/2}{R_{s}-2M-4\pi R_{s}\sigma\sqrt{R_{s}-2M}}.
\end{equation}
For $4\pi R_{s}\sigma<<\sqrt{1-2M/R_{s}}$ the equation of state becomes
approximately linear%
\begin{equation}
P\approx\frac{2M}{R_{s}-2M}\ \sigma
\end{equation}
and a mixed equation of state results as the effect of $M$ increases. \ 

\section{Conclusion}

A general stress-energy-motion relation was derived for Israel layers between
Schwarzschild manifolds. The equation generalizes the equation of state for
layers dropping on exterior Schwarzschild geodesics. It was used to discuss
the relation between motion and equation of state. The motion input is a
single function of the exterior comoving observer acceleration and velocity.
Using the relation, the motion of a layer with equation of state $P=-\sigma/4$
was shown to be linearly related to geodesic motion.

A set of linked L-layers with a common motion was described over the parameter
range $-1\leq a\leq-1/2,$ $0\leq a\leq1/2.$ In the linked range, each layer
with pressure has a partner layer with tension. There is also a set of linked
shells, both with tension in the range $-1/2<a<0.$ These layers coincide for
$P=-\sigma/4$. There are layers in the range $1/2<a\leq1$ but they are not
linked to physical layers with tension. Positive $^{\prime}a^{\prime}$ values
in this range are linked\ to negative values larger than $1$. Because the
motion of the linked shells is the same, their motion functions agree and the
new relation can be used to compare densities and pressures. \ 

The stress-energy in the Israel formalism is described by two observers
comoving with the layer. As pointed out by Ipser and Skivie, \cite{IS84,Ips84}
, the existence of static layers with tension is related to the accelerations
of the comoving observers. For a static layer, the two observers are hovering
over the layer but must accelerate in order to remain static with respect to
the layer. For L-layers, the radially projected 4-accelerations given in
Eq.(\ref{u-diff}) are related to the size of state parameter $^{\prime
}a^{\prime}$.%
\begin{equation}
n_{r+}\dot{U}_{+}^{r}-n_{r-}\dot{U}_{-}^{r}=4\pi\sigma(1+2a).
\end{equation}
This can be used to interpret the relative sizes of $\dot{U}^{r}$ needed for
the two hovering observers. For the $M=0$ case, the interior observer is not
accelerated at all and this equation describes whether the\ projected
4-acceleration needed by the exterior Schwarzschild observer points inward or
outward. For $-1/2<$ $a$ the observer needs to accelerate away from the layer,
counteracting the gravitational attraction of the layer and for $a<-1/2$, the
region where there are static shells with tension, the observer has to
accelerate toward the layer. The repelling nature of layers with tension has
also been discussed by Vilenkin \cite{Vil81} .\ 

The applications focused mainly on L-layers because the $P/\sigma$ structure
of the geodesic extension makes this a simple example to develop with clarity.
\ There are many interesting questions yet to be studied. Other equations of
state easily could be investigated, for example, a dynamic first order
polytrope with $P=K\sigma^{2}$ would have a density%
\[
\sigma(1+4K\sigma)^{3}=(1+\gamma)^{3/2}\left(  F-2K\sigma\right)  ^{2}\left[
\frac{1+F+2K\sigma}{\pi(m_{0}-M)}\right]  .
\]
Using this equation, the suggestion that low density static shells are first
order polytropes could be explored for dynamic shells. The general
stress-energy-motion equation has been developed for Schwarzschild but could
be a useful tool in understanding layers bounding other metrics.\ The layer
linkage discussed here depends on the radial structure of $\Delta$
in\ Schwarzschild L-layers. The idea of linkages for L-layers bounding other
metrics in a variety of dimensions may have broad applications.

\appendix

\section{Observer Acceleration}

The layer is tracked by two observers comoving with the layer who agree on the
layer metric. The radial 4-accelerations of these observers are computed from
the 4-acceleration%
\begin{equation}
A^{i}=U^{a}\nabla_{a}U^{i}=U^{a}\partial_{a}U^{i}+\Gamma_{ab}^{i}U^{a}U^{b}.
\end{equation}
Using%
\[
\frac{d\dot{R}}{d\tau}=\frac{\partial\dot{R}}{\partial t}\frac{\partial
t}{\partial\tau}+\frac{\partial\dot{R}}{\partial r}\frac{\partial r}%
{\partial\tau}=\frac{\partial\dot{R}}{\partial t}\dot{T}+\frac{\partial\dot
{R}}{\partial r}\dot{R}%
\]
one finds%
\begin{equation}
A^{r}=\frac{d\dot{R}}{d\tau}+[\frac{m_{0}}{R^{2}}f](\dot{T})^{2}-[\frac{m_{0}%
}{R^{2}}(1/f)](\dot{R})^{2}.
\end{equation}
The velocity normalization is%
\[
f\dot{T}^{2}=\dot{R}^{2}/f+1.
\]
For the exterior observer we have%
\begin{equation}
A^{r}=\ddot{R}+\frac{m_{0}}{R^{2}}.
\end{equation}
Thus, in general%
\begin{equation}
\dot{U}_{\pm}^{r}=\ddot{R}+(1-f_{\pm})/(2R). \label{gen-accel}%
\end{equation}

\section{$P=a\sigma$}

\subsubsection*{General Motions}

Because of its simplicity, the linear equation of state is often used with the
field equations and it is frequenly used in discussing thin shells
\cite{LW86,GV89,NU201}.\ For layers with $P=a\sigma$, we have%
\begin{equation}
\frac{\dot{\Delta}}{\Delta}=-(1+2a)\frac{\dot{R}}{R}.
\end{equation}
If $^{\prime}a^{\prime}$ is a constant we have%
\begin{align}
\Delta &  =-c_{a}R^{-(1+2a)}.\\
4\pi P_{a}  &  =ac_{a}R^{-2(1+a)}.\nonumber\\
4\pi\sigma_{a}  &  =c_{a}R^{-2(1+a)}.\nonumber\\
m_{L_{a}}  &  =c_{a}R^{-2a}.\nonumber
\end{align}
$c_{a}$ carries an $^{\prime}a^{\prime}\ $index because, from a unit
standpoint, it will have to vary with the value of $^{\prime}a^{\prime}$. The
motion of the layer is described by
\begin{align}
\dot{R}^{2}  &  =\left(  \frac{\Delta}{2}\right)  ^{2}+\left(  \frac{m_{0}%
-M}{R\Delta}\right)  ^{2}+\frac{m_{0}+M}{R}-1\nonumber\\
&  =\frac{c_{a}^{2}}{4R^{2(1+2a)}}+\frac{(m_{0}-M)^{2}R^{4a}}{c_{a}^{2}}%
+\frac{m_{0}+M}{R}-1.\\
\ddot{R}  &  =-\frac{c_{a}^{2}(1+2a)}{4R^{3+4a}}+\frac{2a(m_{0}-M)^{2}%
R^{4a-1}}{c_{a}^{2}}-\frac{m_{0}+M}{2R^{2}}. \label{r-two-dot}%
\end{align}
The points $\dot{R}=0\ $provide an equation for $c_{a}$. \
\begin{equation}
2R_{0}^{1+4a}[R_{0}-m_{0}-M\pm\sqrt{R_{0}^{2}-2R_{0}(m_{0}+M)+4m_{0}M}%
]=c_{a}^{2} \label{ca-2}%
\end{equation}
Not all values of $^{\prime}a^{\prime}$ will correspond to a static layer with
both $\dot{R}=0$ and $\ddot{R}=0.$ Where there is a static layer, the
constant, $c_{a}$, can be evaluated in terms of the equation of state
parameter $^{\prime}a^{\prime}$. \ \ 

\subsubsection*{Static layer}

The static points, $R_{s}$, follow from $\ddot{R}_{s}=0,$ $\dot{R}_{s}%
=0.$\ This identifies $R_{0}$ with $R_{s}$. Using Eqs.(\ref{r-two-dot}%
,\ref{ca-2}) one finds%
\begin{equation}
(1+2a)c_{a}^{4}R_{s}^{-4(1+2a)}-\frac{8a(m_{0}-M)^{2}}{R_{s}^{2}}%
+\frac{2(m_{0}+M)c_{a}^{2}R_{s}^{-2(1+2a)}}{R_{s}}=0.
\end{equation}
The parameter values, $a=0,$ $a=-1/2$ have no solutions. The static radius in
terms of $c_{a}$ is
\begin{equation}
R_{s}^{1+4a}=\frac{-(1+2a)c_{a}^{2}}{(m_{0}+M)\pm\sqrt{(m_{0}+M)^{2}%
+8a(1+2a)(m_{0}-M)^{2}}}.
\end{equation}
Using Eq.(\ref{ca-2}) for $c_{a}$, The static layer radius is%
\begin{equation}
R_{s}=\frac{(m_{0}+M)(1+4a)^{2}\pm\sqrt{(m_{0}+M)^{2}(1+4a)^{4}-32am_{0}%
M(1+2a)(1+4a)^{2}}}{8a(1+2a)}.
\end{equation}
For $M=0$ the layer description is especially simple and we have%
\begin{align}
R_{s}  &  =m_{0}\frac{(4a+1)^{2}}{4a(1+2a)}.\\
c_{a}^{2}  &  =R_{s}^{1+4a}\frac{4am_{0}}{1+2a}.\nonumber
\end{align}%
\begin{align}
4\pi\sigma_{a}  &  =\frac{1}{m_{0}}\frac{(4a)^{2}(1+2a)}{(4a+1)^{3}}.\\
4\pi P_{a}  &  =\frac{1}{m_{0}}\frac{16a^{3}(1+2a)}{(4a+1)^{3}}.\nonumber\\
m_{L_{a}}  &  =m_{0}\frac{1+4a}{(1+2a)}.\nonumber
\end{align}
over the $^{\prime}a^{\prime}$ range%
\begin{align*}
-1  &  \leq a<-1/2\\
0  &  <a\leq1
\end{align*}
It is clear that there are static shells with tension as well as with
pressure. \ 

\section{Derivation of geodesic extension}

The extension is developed in terms of $A$ and $\gamma,$ which are zero for
exterior geodesic motion. \
\begin{align}
A  &  =\ddot{R}+m_{0}/R^{2}.\\
\gamma &  =\dot{R}^{2}-2m_{0}/R.
\end{align}
The Israel formalism gives the stress and density in terms of a function
$\Delta,$%
\begin{align*}
8\pi P  &  =\Delta/R+\dot{\Delta}/\dot{R},\\
4\pi\sigma &  =-\Delta/R,
\end{align*}
with%
\begin{subequations}
\begin{align}
\Delta_{+}  &  =\sqrt{1+\dot{R}^{2}-2m_{0}/R}=\sqrt{1+\gamma}.\\
\Delta_{-}  &  =\sqrt{1+\dot{R}^{2}-2M/R}=\sqrt{1+\gamma+2\frac{m_{0}-M}{R}%
}.\\
\Delta &  =\Delta_{+}-\Delta_{-}%
\end{align}
Calculating the derivatives with respect to $\tau$ we have%
\end{subequations}
\begin{align}
\overset{\cdot}{\Delta}_{+}  &  =\frac{\dot{R}(\ddot{R}+m_{0}/R^{2})}%
{\sqrt{1+\dot{R}^{2}-2m_{0}/R}}=\frac{\dot{R}A}{\Delta_{+}}.\label{dot-plus}\\
\overset{\cdot}{\Delta}_{-}  &  =\frac{\dot{R}(\ddot{R}+M/R^{2})}{\sqrt
{1+\dot{R}^{2}-2M/R}}=\frac{\dot{R}(A+M/R^{2}-m_{0}/R^{2})}{\Delta_{-}}.
\label{dot-minus}%
\end{align}
There are several useful relations. Using Eq.(\ref{dot-plus}) and
(\ref{dot-minus}) one finds%
\begin{equation}
\Delta_{+}^{2}-\Delta_{-}^{2}=2\frac{M-m_{0}}{R}%
\end{equation}
and using this, the density of the layer is%
\begin{equation}
4\pi\sigma=\frac{(\Delta_{+}-\Delta_{-})(\Delta_{+}^{2}-\Delta_{-}^{2}%
)}{2(m_{0}-M)}.
\end{equation}
A pressure-density relation can also be found%
\begin{align}
8\pi PR  &  =-4\pi\sigma R+\frac{RA}{\Delta_{+}}-\left(  A+\frac{M-m_{0}%
}{R^{2}}\right)  \frac{R}{\Delta_{-}}.\nonumber\\
8\pi PR+4\pi\sigma R  &  =\frac{RA(\Delta_{-}-\Delta_{+})}{\Delta_{-}%
\Delta_{+}}-\frac{M-m_{0}}{R\Delta_{-}},\nonumber\\
8\pi PR+4\pi\sigma R  &  =\frac{RA(\Delta_{-}-\Delta_{+})}{\Delta_{-}%
\Delta_{+}}-\frac{\Delta_{+}^{2}-\Delta_{-}^{2}}{2\Delta_{-}},\nonumber\\
8\pi PR+4\pi\sigma R  &  =\frac{RA(4\pi\sigma R)}{\Delta_{-}\Delta_{+}}%
+\frac{(\Delta_{+}+\Delta_{-})4\pi\sigma R}{2\Delta_{-}}.\nonumber\\
2P/\sigma+1/2  &  =\frac{AR}{\Delta_{-}\Delta_{+}}+\frac{\Delta_{+}}%
{2\Delta_{-}}.\nonumber\\
(4P/\sigma+1)\Delta_{-}  &  =\frac{2AR}{\Delta_{+}}+\Delta_{+}\ .
\label{p-over-sig}%
\end{align}
A useful relation is%
\begin{equation}
(4\frac{P}{\sigma}+1)\frac{\Delta_{-}}{\Delta_{+}}=\frac{2AR}{1+\gamma}+1,
\end{equation}
with%
\begin{subequations}
\begin{align}
\frac{\Delta_{-}}{\Delta_{+}}+1  &  =\frac{2AR/(1+\gamma)+2+4P/\sigma
}{(4P/\sigma+1)},\\
\frac{\Delta_{-}}{\Delta_{+}}-1  &  =\frac{2AR/(1+\gamma)-4P/\sigma
}{(4P/\sigma+1)}.
\end{align}
Using these, substituting into Eq.(\ref{p-over-sig}), the general extension of
the geodesic equation of state follows.%
\end{subequations}
\begin{equation}
\pi(m_{0}-M)\sigma(4\frac{P}{\sigma}+1)^{3}=(1+\gamma)^{3/2}\left[
F-2P/\sigma\right]  ^{2}\left[  F+1+2P/\sigma\right]  .\
\end{equation}
For$\ A=\gamma=0,$ this becomes the geodesic equation of state.


\begin{thebibliography}{99}                                                                                               %


\bibitem {Isr66}W. Israel, Nuov. Cim. \textbf{44B}, 1 (1966), Nuov. Cim.
\textbf{48B}, 463\ (1967). \emph{Singular Hypersurfaces and Thin Shells in
General Relativity}

\bibitem {KL79}K. Lake, Phys. Rev. D \textbf{19}, 2847 (1979). \emph{Thin
Spherical Shells}

\bibitem {KM91}M. Khorrami and R. Mansouri, Phys. Rev. D \textbf{44}, 557
(1991). \emph{Spherically symmetric thin walls}

\bibitem {MK96}R. Mansouri and M. Khorrami, J. Math. Phys. \textbf{37}, 5672
(1996). \emph{The equivalence of Darmois-Israel and distributional method for
thin shells in general relativity}

\bibitem {Lob06}F.S.N. Lobo, Phys. Rev. D\textbf{\ 75, }024023 (2007).
\emph{Van der Waals quintessence stars}

\bibitem {Gon02}S.M.C.V. Gon\c{c}alves, Phys. Rev, D \textbf{66}, 084021-1
(2002). \emph{Relativistic shells: Dynamics, horizons, and shell crossing}

\bibitem {LW86}K. Lake and R. Wevrick, Can. J. Phys. \textbf{64}, 165
(1986).\textbf{ }\emph{Evolution of Bubbles in Vacuum }

\bibitem {L-CM86}P. Laguna-Castillo and R.A. Matzner, Phys. Rev. D
\textbf{34}, 2913 (1986). \emph{Inflation and Bubbles in General Relativity}

\bibitem {BKT87}V.A. Berezin, V.A. Kuzmin and I.I. Trachev, Phys. Rev. D
\textbf{36}, 2919 (1987). \emph{Dynamics of Bubbles in General Relativity}

\bibitem {Ips87}J.R. Ipser, Phys. Rev. D \textbf{36}, 1933 (1987).
\emph{Double-Bubble Spacetimes}

\bibitem {KMM06}J. Kijowski, G. Magli and D. Malafarina, Gen. Rel. Gravit.
\textbf{38}, 1697 (2006). \emph{Relativistic dynamics of spherical timelike
shells}

\bibitem {IS84}J.R. Ipser and P. Sikivie, Phys. Rev. D \textbf{30}, 712
(1984). \emph{Gravitationally repulsive domain wall}

\bibitem {Ips84}J.R. Ipser, Phys. Rev. D \textbf{30}, 2452 (1984).
\emph{Repulsive and attractive planar walls in general relativity}

\bibitem {GV89}D. Garfinkle and C. Vuille, Class. Quan. Grav. \textbf{6}, 1819
(1989). \emph{Thin walls in regions with vacuum energy}

\bibitem {MM01}P.O. Mazur and E. Mottola, arXiv/gr-qc/0109035
\emph{Gravitational condenstate stars: an alternative to black holes}

\bibitem {VW04}M. Visser and D.L. Wiltshsire, Class. Quan. Grav. \textbf{21},
1135 (2004). \emph{Stable gravastars-an alternative to black holes. }

\bibitem {Car05}B.M.N. Carter, Class. Quan. Grav. \textbf{22}, 4551
(2005).\emph{ Stable gravastars with generalized exteriors}

\bibitem {CFV05}C. Cattoen, T. Faber and M. Visser, Class. Quan. Grav.
\textbf{20}, 4189 (2005). \emph{Gravastars must have anisotropic pressures.}

\bibitem {deBHI+06}A. deBenedictis, D. Horvat, S. Ilijic, S. Kloster and
K.S.Viswanathan, Class. Quan. Grav. \textbf{23}, 2303 (2006). \emph{Gravastar
solutions with continuous pressures and equation of state. }

\bibitem {ES05}E.F. Eiroa and C. Simeone, Phys. Rev. D \textbf{71, }127501
(2005). \emph{Thin shell wormholes in dilaton gravity}

\bibitem {BI05}C. Barrabes and W. Israel, Phys. Rev. D \textbf{71, }064008
(2005). \emph{Lagrangian brane dynamics in general relativity and
Einstein-Gauss-Bonnet gravity}

\bibitem {GW07}E. Gravanis and S. Willison, Phys. Rev. D \textbf{75,
}084025\textbf{\ }(2007). \emph{Mass without mass: from thin shells in
Gauss-Bonnet gravity}

\bibitem {Bar91}C. Barrabes and W. Israel, Phys. Rev. D \textbf{43}, 1129
(1991). \emph{Thin shells in general relativity and cosmology: The lightlike
limit}

\bibitem {Poi04}E. Poisson, \textit{A Relativist's Toolkit} (Cambridge
University Press, Cambridge 2004).

\bibitem {CI68}V. de la Cruz and W. Israel, Phys. Rev. \textbf{170}, 1187
(1968). \emph{Spinning Shell as a Source of the\ Kerr metric}

\bibitem {ML97}P. Musgrave and K. Lake, Class. Quan. Grav. \textbf{13}, 1885
(1996). \emph{Junctions and thin shells in general relativity using computer
algebra: \ I. The Darmois-Israel formalism. }

\bibitem {Ans02}S. Ansoldi, Class. Quan. Grav. \textbf{19}, 6321 (2002).
\emph{WKB Metastable quantum states of a de Sitter-Reisner Nordstrom dust
shell}

\bibitem {BLP91}P.R. Brady, J. Louko and E. Poisson, Phys. Rev. D \textbf{44},
1891 (1991). \emph{Stability of a shell around a black hole. }

\bibitem {Ver02}R. Vera, Class. Quan. Grav. \textbf{19}, 5249 (2002).
\emph{Symmetry-preservings matchings}

\bibitem {LC05}F.S.N. Lobo and P. Crawford, Class. Quan. Grav. \textbf{22}%
,.4869 (2005). \emph{Stability Analysis of dynamic thin shells}

\bibitem {Mar05}M. Mars,\ Class. Quan. Grav. \textbf{22}, 3325 (2005).
\emph{First and second-order perturbations of hypersurfaces}

\bibitem {KG07}J.P. Krisch and E.N. Glass, Phys. Rev. D \textbf{76}, 104006
(2007). \emph{Collapsing layer on Schwarzschild-Lemaitre geodesics}

\bibitem {NU201}U.S. Nilsson and C. Uggla, Annals Phys. \textbf{286},
278\ (2001) (gr-qc/0002021) \emph{General Relativistc Stars: \ Linear
Equations of State}

\bibitem {Vil81}A. Vilenkin, Phys. Rev. \textbf{D 23}, 852 (1981).
\emph{Gravitational Field of Vacuum Domain Walls and Strings.}
\end{thebibliography}
\end{document}